\documentclass[twocolumn,showpacs,preprintnumbers,amsmath,amssymb,nofootinbib]{revtex4}
\usepackage{graphicx}
\usepackage{dcolumn}
\usepackage{bm}

\global\long\def\ave#1{\left\langle #1 \right\rangle }

\begin{document}

\title{Hydrodynamics at large baryon densities: Understanding proton vs. anti-proton $v_2$ and other puzzles}

\author{J.~Steinheimer}
 \email{jsfroschauer@lbl.gov}
\affiliation{Lawrence Berkeley National Laboratory, 1 Cyclotron Road, Berkeley, CA 94720, USA}

\author{V.~Koch}
\affiliation{Lawrence Berkeley National Laboratory, 1 Cyclotron Road, Berkeley, CA 94720, USA}

\author{M.~Bleicher}
\affiliation{Institut f\"ur Theoretische Physik, Goethe-Universit\"at, Max-von-Laue-Str.~1,
D-60438 Frankfurt am Main, Germany}
\affiliation{Frankfurt Institute for Advanced Studies (FIAS), Ruth-Moufang-Str.~1, D-60438 Frankfurt am Main,
Germany}

\date{\today}

\begin{abstract}
We study the importance of the initial state,
baryon stopping and baryon number transport for the dynamical evolution
of a strongly interacting system produced in heavy ion collisions. 
We employ a hybrid model, which combines the fluid dynamical evolution of
the fireball with a transport treatment for the initial state and the
final hadronic phase. We present results for collisions at beam
energies from $\sqrt{s_{NN}}=7.7$ to $200$ GeV. We study various
observables such as the centrality dependent freeze out
parameters, the non-monotonic behavior of effective slope parameter
parameter with particle mass as well as the apparent difference in
particle and anti-particle elliptic flow. Our results are in
reasonable agreement with the available data. We find that the
propagation of the baryon-number current 
in the hydrodynamic evolution as well as
the transport treatment of the hadronic phase are essential for
reproducing the experimental data.
\end{abstract}

\maketitle
\section{Introduction}
Over the last decades many experimental programs at the
Brookhaven National Laboratory and CERN facilities have 
been devoted to finding signals  of a
new state of matter, the Quark Gluon Plasma (QGP), by means of
relativistic heavy ion collisions. These experiments have produced a 
wealth of data including particle
ratios and yields, transverse and longitudinal momentum spectra as well as the
coefficients of a Fourier decomposition of the transverse flow
patterns
\cite{Ollitrault:1992bk,Rischke:1996nq,Sorge:1996pc,Heiselberg:1998es,Scherer:1999qq,Soff:1999yg,Brachmann:1999xt,Csernai:1999nf,Zhang:1999rs,Kolb:2000sd,Bleicher:2000sx,Stoecker:2004qu,Zhu:2005qa,Petersen:2006vm,Gazdzicki:2004ef,Gazdzicki:1998vd}. Several observations at the Relativistic Heavy Ion
Collider (RHIC) and Large Hadron collider (LHC) indicate that a strongly
coupled QGP (sQGP) dominates the dynamical evolution
\cite{Adams:2005dq,Back:2004je,Arsene:2004fa,Adcox:2004mh,Aamodt:2010cz,Aamodt:2010jd,Aamodt:2010pa,Aamodt:2010pb}. In addition the energy dependencies of various observables, such as the $K/\pi$ ratio,
already show anomalies at low SPS energies which might
be related to the onset of deconfinement and chiral symmetry
restoration at lower energies
\cite{Gazdzicki:2004ef,Gazdzicki:1998vd}. While these are intriguing
observations, experience has taught us that the interpretation of
experimental results and their relation to the deconfinement phase
transition is often ambiguous and extensive model studies are required
to understand the numerous observables.

In order to study the physics of heavy ion collisions many models
have been developed which address specific aspects of the these
reactions. For example, hadron ratios are well described by thermal
models which employ a hadronic resonance gas at a fixed temperature
and chemical potential for the description of particle
yields. The study of transverse and elliptic flow observables require
more complex models: at the lowest energies hadronic transport,
including the effects of hadronic potentials/interactions ( see
e.g. \cite{Bass:1998ca,Bleicher:1999xi,Geiss:1998ki,Cassing:1999es,Petersen:2006vm,Bleicher:2005ti})
are applied, while at the highest RHIC and LHC energies a description of the
system in terms of fluid dynamics seems to be successful \cite{Bass:2000ib,Teaney:2001av,Hirano:2005xf,Nonaka:2006yn,Romatschke:2007mq,Hirano:2010je,Schenke:2011tv,Petersen:2011sb}. 
Since it is desirable to obtain a more comprehensive picture of the
whole dynamics of heavy ion reactions, various so called hybrid
approaches have been developed during the last years
\cite{Magas:2001mr}. In these models one commonly uses initial
conditions that are calculated in a non equilibrium model which are
followed by an ideal or viscous hydrodynamic evolution. For the late stage of the collision a kinetic approach is more appropriate and, therefore, subsequent to the hydrodynamic evolution a transport model solving the Boltzmann equation is used for the description of the freeze-out stage \cite{Paiva:1996nv,Aguiar:2001ac,Socolowski:2004hw,Hirano:2005xf,Hirano:2007ei,Bass:1999tu,Nonaka:2006yn,Dumitru:1999sf,Schenke:2010nt}. Alternative
approaches couple a partonic phase to a hadronic transport model
\cite{Lin:2004en,Cassing:2009vt,Cassing:2008sv}. As each phase may contribute considerably to final state observables, the interpretation of experimental results and their relation to the deconfinement phase transition are often a difficult task, and the contributions of the different phases need to be evaluated thoroughly.\\

The purpose of this paper is to study heavy ion collisions within
  the framework of one such hybrid model, the so called UrQMD hybrid
  model \cite{Petersen:2008dd} with the aim to re-evaluate some of the commonly
  accepted interpretations of various observables. Specifically we 
will address the centrality dependence of freeze out parameters, transverse momentum spectra as well as anti-particle elliptic flow at low energies.

This paper is organized as follows: First we briefly introduce the
hybrid model. Next we discuss particle production and transverse
spectra obtained with the model. In the final section we present our
results for the elliptic flow of particles and anti-particles with an
emphasis on the difference between proton and anti-proton flow.

\section{The hybrid model}\label{s2}
The UrQMD hybrid model combines the advantages of a hadronic transport model with an intermediate hydrodynamical stage for the hot and dense phase of a heavy ion collision. The UrQMD Model \cite{Bass:1998ca,Bleicher:1999xi} (in its cascade mode) is used to calculate the initial state of a heavy ion collision for the hydrodynamical evolution \cite{Steinheimer:2007iy}. This is done to account for the non-equilibrium dynamics in the very early stage of the collision. The coupling between the UrQMD initial state and the hydrodynamical evolution happens at a time $t_{\rm start}$ when the two Lorentz-contracted nuclei have passed through each other.

\begin{equation}
t_{start} = \frac{2R}{\sqrt{\gamma_{C.M.}^2 -1}}	
\end{equation}

where $\gamma_{C.M.}$ is the center of mass frame Lorentz factor and $R$ is the radius of the nucleus.
At this starting time all initial collisions have happened. It is further the earliest time at which local thermodynamical equilibrium may be achieved.
At this time the energy, baryon number and momenta of all particles within UrQMD are mapped onto the spatial grid of the hydrodynamic model by representing each hadron by a Gaussian of finite width $\sigma=1$~fm. In this 
approach the effects of event-by-event fluctuations and stopping of energy and baryon number density in the initial state are naturally included.

The full (3+1) dimensional, one fluid, ideal hydrodynamic evolution is performed using the SHASTA algorithm \cite{Rischke:1995ir,Rischke:1995mt}. We solve the equations for the conservation of energy and momentum:

\begin{equation}
	\partial_{\mu}T^{\mu \nu}= 0
\end{equation}
and for the conservation of the baryonic current,
\begin{equation}\label{densprop}
	\partial_{\mu} N^{\mu} = 0
\end{equation}
Here $T^{\mu \nu}$ is the relativistic energy momentum tensor,
\begin{equation}
	T^{\mu \nu}=(\epsilon + p)u^{\mu}u^{\nu}-g^{\mu \nu} p
\end{equation}
and $ N^{\mu}$ the baryonic current
\begin{equation}
	 N^{\mu}=n \ u^{\mu}	 
\end{equation}
 
The above partial differential equations are solved on a three-dimensional spatial Eulerian grid with fixed position and size $\delta x = 0.2$ fm in the computational frame. The local rest frame is defined as the frame where $T^{\mu \nu}$ has diagonal
form (i.e. all off-diagonal elements vanish), also known as the Landau
frame. To close the set of equations an equation of state (EoS), the pressure as function of energy and baryon number density $p(\epsilon,n)$ needs to be specified. In the following we will use an EoS that corresponds to an hadron resonance gas. This is convenient as it includes the same degrees of freedom as the UrQMD model, which essentially allows us to study the effect of local equilibrium on the different observables. Previous investigations have shown that different equations of state only lead to insignificant differences in the results \cite{Petersen:2009mz,Steinheimer:2009nn}.

To transfer all particles back into the UrQMD model, an approximate
iso-eigentime transition is chosen (see \cite{Li:2008qm} for
details). To this end we apply the Cooper Frye prescription
\cite{Cooper:1974mv} to individual transverse slices, of thickness
$\Delta z = 0.2 $fm, at a time-like transition
hypersurface. The transition time for a given slice is determined
by the time when the energy density $\varepsilon$ in every cell of this slice
has dropped below five times the nuclear ground state energy density,
i.e. below $\sim 730 {\rm
MeV/fm}^3$. As a result we obtain a longitudinal iso-eigentime
transition with an almost rapidity independent maximum switching temperature
for beam energies above $\sqrt{s_{NN}} \approx 10$ GeV. \footnote{This ensures that all
cells have passed through a potential mixed phase of the equation of state and
the effective degrees of freedom at the transition are hadronic. As mentioned above we use a hadron resonance gas EoS
for our present study which does not include a phase transition. However, we could include an EoS with a phase transition, even though most bulk observables have shown to be insensitive on the specific choice of the EoS \cite{Steinheimer:2009nn,Li:2008qm,Petersen:2009mz}.} In a
given slice the hydrodynamic fields are transformed to particle degrees of freedom via the Cooper-Frye equation:

\begin{equation}\label{cooper}
	E \ \frac{dN}{d^3p} = g_i \int_{\sigma}{f_i(x,p) \ p^{\mu} \ d\sigma_{\mu}}
\end{equation}
on an isochronous time-like hypersurface in the computational frame
(the hypersurface normal is $d\sigma_{\mu}=(d^3x,0,0,0)$). Here
$f_i(x,p)$ are the distribution functions and $g_i$ the degeneracy
factors for the different particle species $i$, which in our case are
given by the appropriate Bose- or Fermi-distributions. The particle distributions explicitly depend on the local values of the temperature $T$ and baryon chemical potential $\mu_B$. $T$ and $\mu_B$ are obtained by converting the local energy- and baryon-densities via the equation of state.\\

The above transition procedure  conserves
baryon number, electric charge, the total net strangeness and the
total energy on an event-by-event basis. 
After the particles are created they
evolve according to a hadronic cascade (UrQMD) where final
re-scatterings and decays are calculated until all interactions cease. 
A more detailed description of the hybrid model
including parameter tests and results can be found in
\cite{Petersen:2008dd}.

\subsection{Particle yields and the Temperature distribution}

The description of particle production in heavy ion collisions
using statistical methods, such as thermal models
\cite{Becattini:2003wp,Andronic:2008gu,Cleymans:2005xv} and fluid
dynamical models with a sudden freeze out, are usually based on the
assumption that for a given incident beam energy chemical freeze out occurs at a fixed temperature and
chemical potential. As we have discussed above (see also
\cite{Steinheimer:2009nn}) we transform full transverse slices at a
given time $t$, using the Cooper Frye prescription. This implies that
contrary to an isothermal transition particles are emitted over a range of temperatures/densities.
While this may no be a standard procedure it has
been shown in several publications
\cite{Petersen:2008dd,Petersen:2008gy,Steinheimer:2010zza,Santini:2011zw,Steinheimer:2011mp,Steinheimer:2012tb}
that this hybrid model is very well suited to describe particle ratios and yields over a wide range of energies.
In addition, recent comparisons of an iso-thermal 3+1 dimensional hypersurface \cite{Huovinen:2012is} indicate that deviations only occur at large transverse
momenta ($p_{\bot}>1$ GeV), corresponding in our freeze out
prescription to very cold and fast cells which are
being treated as equilibrated for an extended period of time. Since
all observables we discuss in this paper are mainly determined at
lower momenta our results hold also for the iso-thermal case. The comparison also implies that
both dynamical descriptions, transport and hydrodynamics, give equivalent results for the fireballs
expansion in the transition region considered.

 \begin{figure}[t]
 \centering
\includegraphics[width=0.5\textwidth]{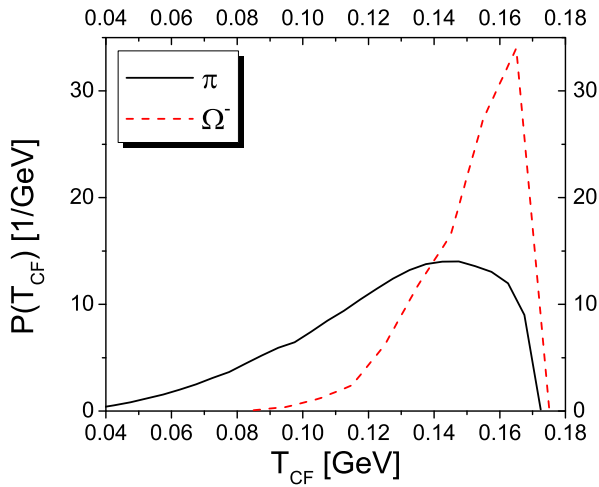}
 \caption{(Color online) Probability distributions $P(T_{CF})$ for pions
   (black solid line) and $\Omega^-$'s (red dashed line) to be
   produced from a cell with a give temperature $T_{CF}$. The results shown
   are for central collisions of Pb nuclei with a beam energy of
   $\sqrt{s_{NN}}= 17.3$ GeV.}
 \label{tdistr}
\end{figure}
 
\begin{figure}[t]
 \centering
\includegraphics[width=0.5\textwidth]{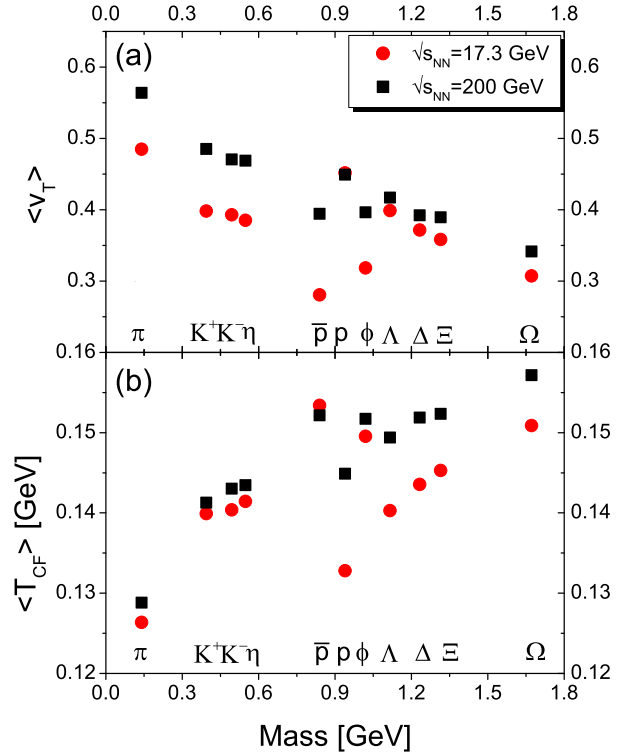}
 \caption{(Color online) The average transverse flow velocity $\left< v_{\bot} \right>$ (figure (a)) and  $T_{CF}$ (figure (b)) at the Cooper Frye transition, as a function of the particle mass for different hadron species. Results for most central Pb+Pb/Au+Au collisions at $\sqrt{s_{NN}} = 17.3$ (black squares) and $200$ GeV (red circles) are shown. Note the anti-proton mass has been shifted for visibility.}
 \label{3}
\end{figure}

Figure \ref{tdistr} shows the resulting probability distributions
$P(T_{CF})$ for pions and $\Omega^-$'s to be produced from a cell with a
given temperature $T_{CF}$. Note that $T_{CF}$ is the temperature that enters into the Cooper-Frye equation (\ref{cooper}). The results shown are for central collisions of
Pb nuclei at a center of mass energy of $\sqrt{s_{NN}} = 17.3$~GeV. One can clearly see that the pions, due to their smaller mass, are generally emitted/produced
at lower temperatures than the $\Omega$'s. In other words, because we allow for emission
from all local temperatures the heavier particles are preferentially
emitted from regions of higher temperatures. 
The lower plot of Fig.~\ref{3} illustrates how the average value of $T_{CF}$ changes as a function of the particle mass for different hadron species. Results for Pb+Pb/Au+Au collisions at $\sqrt{s_{NN}} = 17.3$ and $200$~GeV are shown. One observes a clear trend of increasing $T_{CF}$ with the particle mass, as shown explicitly for the pion and $\Omega$. 

We also observe a mass dependence of
the average transverse flow $\left< v_{\bot} \right>$, defined as the average of the hydrodynamical radial flow velocity over the production points of the particles. 
Results for the different particle species are depicted in the upper part of figure \ref{3}. It is clear that hot and dense
cells usually have a smaller flow velocity because they tend to be located
at the center of the collision zone. The colder cells are more
peripheral and therefore have acquired more flow.\\ 

\begin{figure}[t]
 \centering
\includegraphics[width=0.5\textwidth]{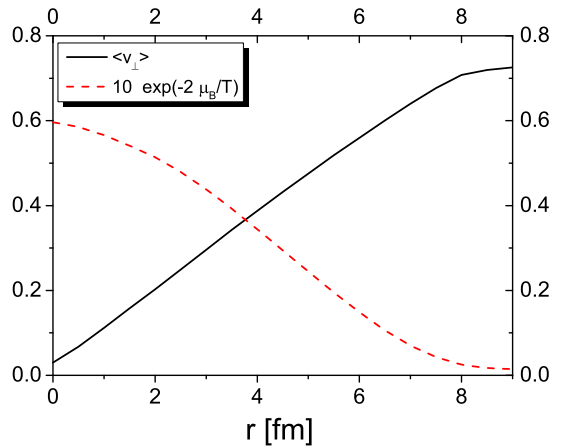}
 \caption{(Color online) Average transverse hydrodynamical flow velocity (black solid line) and the scaled ratio of protons over anti-protons $\approx \exp{(-2 \mu_B/T)}$ (red dashed line) as a function of transverse radius. The results shown are for the central transverse plane at the end of a hydrodynamical calculation for Au+Au at $\sqrt{s_{NN}}=11.5$ GeV.}
 \label{12}
\end{figure}

At lower beam energies, the finite baryon density (chemical potential)
plays an important role. Because the proton has a large chemical
potential, it is much more abundant at lower temperatures than it's
anti-particle, resulting in a distinctly different value of $\left< T_{CF} \right>$, defined as the average of temperatures over the production points of the particular particles. Figure \ref{12} depicts the averaged ratio of
anti-protons over protons, taken at the last time step of an Au+Au
collision at a beam energy of $\sqrt{s_{NN}}= 11.5$ GeV, $\bar{p}/p=\exp{(-2 \ \mu_B/T)}$, as a function of radius
(red dashed line). Clearly the $\bar{p}/p$ ratio is largest in
the center of the collision zone where the radial flow has its
minimum. The transverse flow velocity $v_{\bot}$ of the hydrodynamic fluid (black solid line), extracted from our hydrodynamical calculation, increases linearly
with the distance from the center of the collision. Consequently the
anti-protons will acquire a smaller average transverse flow as
compared to the protons, even though their transverse velocity for any given $r$ is identical. The conclusion that protons and anti-protons acquire a different average
flow velocity also holds true for an
iso-thermal transition because the ratio $\bar{p}/p$ will in
general not be constant over the hypersurface. The quantitative difference however might very well depend on 
the definition of the transition hypersurface. It is therefore interesting to extend the current investigation 
to different hypersurfaces \footnote{For these further studies we need to apply a hypersurface finder for our full 3+1D calculation including local density fluctuations. Such a tool has recently been developed \cite{Huovinen:2012is} and can be applied for future studies}.\\
Note that the same
argument should also hold for other conserved charges, such as
strangeness and the third component of the isospin. Because the
colliding nuclei have more neutrons than protons the third component
of the isospin is finite, and {\em negative} in the produced
fireball. Therefore, we expect the ratio $\pi^+/\pi^-$ to depend on
the radius similar to the $\bar{p}/p$-ratio, which in consequence will
lead to a different transverse
flow velocity of $\pi^+$ compared to $\pi^-$. However the effect will be much weaker for pions since it
depends on the ratio of the fugacities $\exp{(-2 \ \mu_I/T)}$ and the
isospin chemical potential is considerably smaller then the baryon
chemical potential $-\mu_I<\mu_S<\mu_B$. For the top SPS energy a thermal model analysis found values of: $\mu_I = -5.0$ MeV, $\mu_S=71.1$ MeV and $\mu_B=266 \pm 5$ MeV \cite{BraunMunzinger:1999qy}.

\subsection{Centrality Dependence of Freeze Out Parameters}

An ideal fluid dynamical treatment of the expansion implies that for a given
freeze out criterion the particle abundances are fixed by the total entropy per
baryon ($S/A$) produced in the very early stage of the collision, since
the subsequent expansion is isentropic.
  
In all models used to describe the properties of the early stage of an
heavy ion collision (Glauber model, geometrical overlap, hadronic
transport) the initial state is defined by the total energy and baryon
number deposited in the fireball.
In a Glauber model \cite{Ollitrault:1992bk,Blaizot:1990zd,Kolb:2000sd} 
the energy deposition in the transverse plane is proportional to the number of wounded nucleons:
 \begin{eqnarray}
   && e(x,y;\tau_0) =
 \\
   && K \left\{ T_A\bigl(x{+}{\textstyle{b\over 2}},y\bigr)
      \Bigl[1-\Bigl(1-{\sigma T_B\bigl(x{-}{\textstyle{b\over 2}},y\bigr)
                       \over B}\Bigr)^B \Bigr] \right. 
 \nonumber\\
   &&\ \ + \left. T_B\bigl(x{-}{\textstyle{b\over 2}},y\bigr)
     \Bigl[1-\Bigl(1-{\sigma T_A\bigl(x{+}{\textstyle{b\over 2}},y)
                      \over A}\Bigr)^A \Bigr]\right\}\,.
 \nonumber
 \end{eqnarray}
where  $T_A$ and $T_B$ are the nuclear thickness functions of the incoming 
nuclei $A$ and $B$, $\sigma$ is an energy dependent cross section, and $K$ is a proportionality factor. 
If we further assume that the initial baryon density is proportional to
the initial energy density \cite{Kolb:2000sd}, 
 \begin{eqnarray}
   n(x,y;\tau_0) = C(\sqrt{s_{NN}}) \cdot e(x,y;\tau_0)\,,
 \end{eqnarray}
the initial energy per baryon $(E/A)(r)$ as a function of the
  transverse radius is a constant which only depends on the
  collision energy.
At the same time the
energy and baryon densities drop rather quickly as a function of
radius $r$. Since $E/A$ is constant, a decreasing
energy density implies that the entropy per baryon $S/A$ must vary as a function of $r$ as the
entropy density in general does not scale with the energy density,
 \begin{eqnarray}
 {s(x,y;\tau_0) \over e(x,y;\tau_0)} \ne {\rm constant}. 
 \end{eqnarray}
In addition the average energy density varies as a function of
centrality. Since $E/A$ is constant we also expect a centrality
dependence of the total entropy per baryon.\\
 
\begin{figure}[t]
 \centering
\includegraphics[width=0.5\textwidth]{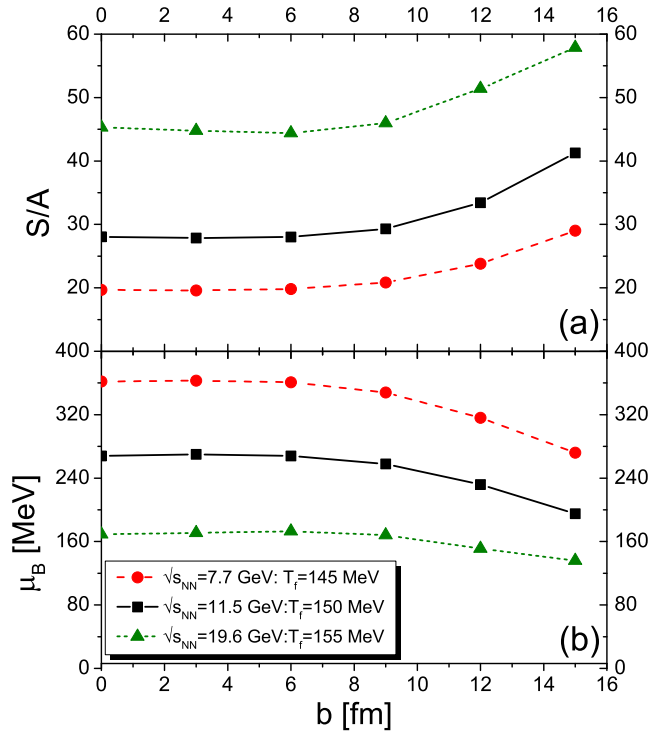}
 \caption{(Color online) (a): Centrality dependence of the Entropy per baryon produced.
 (b): Corresponding baryon chemical potential for a given fixed temperatures $T_f$.}
 \label{1}
\end{figure}

Figure \ref{1} exemplifies the effects of the initial geometry on the
final chemical composition of the fireball. Figure
\ref{1}a shows the total entropy per baryon produced in collisions of
Au nuclei at energies of $\sqrt{(s_{NN})}=7.7$-$19.6$ GeV, as a
function of impact parameter $b$. 
The entropy per baryon is calculated, averaging over 1000 UrQMD initial state events (although it
has been shown that $S/A$ does only vary weakly on an event-by-event basis \cite{Steinheimer:2007iy}).
A Hadron Resonance Gas (HRG)
equation of state is used to estimate the values for the entropy per baryon as it best 
resembles the active degrees of freedom at the particle freeze out. We find that the produced
$S/A$ increases with the impact parameter, simply because of the different
initial geometry. Using the HRG we can determine the baryonic chemical potential
$\mu_B$, corresponding the value of $S/A$ when fixing the temperature to any given value 
$T_f$. For the thermal models of particle production
this temperature corresponds to the chemical freeze out temperature. The resulting values for $\mu_{B}$ are shown in figure \ref{1}b as a function of $b$. It is clear that $\mu_B$ decreases with
increasing $b$. We note that such an effect has been observed in
experiment \cite{Kumar:2012xa}, where thermal fits to different centrality
selections show a decreasing $\mu_B$ with centrality which is of comparable magnitude to our results shown here.
This observation indicates that the systems created in heavy ion
collisions of fixed energy but varying centrality cannot be
characterized by single values for the  thermal parameters.

\section{The slope parameter}

After discussing the integrated yields, let us next discuss the
transverse momentum spectra, which are considered to be more sensitive
to the kinetic decoupling temperature, $T_{th}$, and transverse flow than
average yields. Taking Eq. (\ref{cooper}) and neglecting any flow
the transverse momentum distribution  (for $p_z \rightarrow 0$) is given by

\begin{equation}
	\frac{1}{m_{\bot}} \frac{dN_i}{d m_{\bot} dy} \propto m_{\bot} (\exp{((m_{\bot}-\mu_i)/T)}\pm 1)^{-1}
\end{equation}
with the transverse mass $m_{\bot}=\sqrt{(m_i^2 + p_{\bot}^2)}$. 
The negative slope of $\log(\frac{1}{m_{\bot}^2} \frac{dN_i}{d m_{\bot}dy})$ gives the so called {\em effective} temperature $T_{eff}$
  \cite{Sollfrank:1990qz,Xu:2001zj}, which, however, does not
  correspond to the actual kinetic decoupling temperature $T_{th}$ since the spectra are blue
  shifted due to transverse flow. To account for the flow the
  effective temperature can be related to the kinetic decoupling temperature by $T_{eff}= T_{th} + m_i
  \left< v_{\bot}\right>^2$.
  
These type of  fits to experimental data typically result in a
non-monotonic behavior in $T_{eff}$ as a function of particle mass,
which cannot be accounted for with a fixed temperature freeze
out \cite{Bass:2000ib,Arbex:2001vx,Dumitru:2005hr}. Note that in our calculation we cannot directly extract a value for the kinetic freeze out temperature $T_{th}$, because we treat the kinetic decoupling as a dynamical process within the UrQMD transport model.

In addition to flow effects resonance decays as well as scattering
processes in the hadronic phase affect the extracted value of
$T_{eff}$ as a function of mass \cite{Sollfrank:1990qz,Brown:1991en,Florkowski:2001fp,Bass:2000ib}.  
To successfully interpret the experimental results one, therefore, has to disentangle the different contributions to the transverse spectra in a consistent approach. 

\begin{figure}[t]
 \centering
\includegraphics[width=0.5\textwidth]{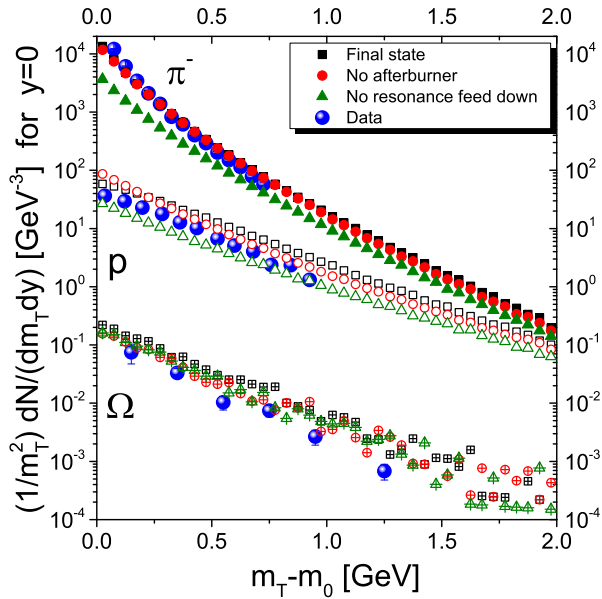}
 \caption{(Color online) Transverse mass distributions of pions protons and $\Omega$'s at mid rapidity $|y|<0.5$, divided by $m_{\bot}^2$, for most central collisions of Pb nuclei at $\sqrt{s_{NN}}= 17.3$ GeV. We compare the distributions from the hybrid model extracted at the final state (black squares) with those directly at the Cooper Frye transition with (red circles) and without (green triangles) resonance decays. Data from the NA49 experiment are depicted as blue circles \cite{Afanasiev:2002mx,Alt:2006dk,Alt:2004kq}.}
 \label{4}
\end{figure}

Using the UrQMD hybrid model we can disentangle all the
important contributions and explore what information about the hot and dense
phase can be extracted from the spectra. Figure \ref{4} shows the mid
rapidity ($|y|<0.5$) $m_{\bot}$ spectra of pions protons and
$\Omega$'s, divided by $m_{\bot}^2$, for the most central collisions of Pb
nuclei at $\sqrt{(s_{NN})}= 17.3$ GeV. We compare the final spectra
(black squares) with the ones obtained directly after the Cooper Frye
transition, without any UrQMD final state, where we either let all resonances decay at the transition (red
circles labeled as 'no afterburner') or do not allow for resonance decays at the transition.
(green triangles labeled as 'no resonance feed down'). The two latter cases are interesting to distinguish because they show the effect of the final state rescattering 
on the momentum distribution functions, indicating that the final hadronic state is not a mere decay of resonances.
For comparison we also show the experimental
data by the NA49 collaboration \cite{Afanasiev:2002mx,Alt:2006dk,Alt:2004kq} as blue circles.

\begin{figure}[t]
 \centering
\includegraphics[width=0.5\textwidth]{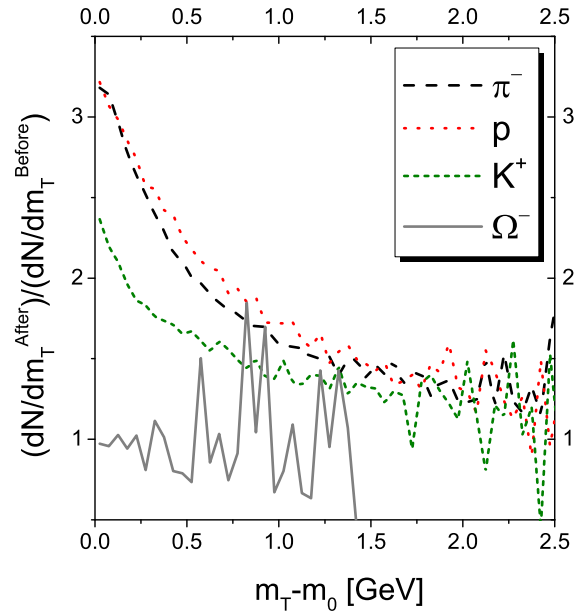}
 \caption{(Color online) Ratios of transverse mass distributions after to before resonance decays, shown for pions protons kaons and $\Omega$'s, as a function of $m_{\bot}$. }
 \label{5}
\end{figure}

We see that the spectra of both pions and protons are significantly
modified due to hadronic interactions and resonance decays. 
We further find deviations of the spectra from an
exponential shape already at the Cooper Frye transition, when no
resonance decays are taken into account. This is a result of
the blue shift due to the finite flow already present at the transition, as shown in Fig.~\ref{3}, and 
is also connected to our freeze out treatment of summing up thermal distributions with different
temperatures, which was first discussed in \cite{Csernai:2005ht,Molnar:2005gx,Molnar:2005gy}.

We note that similar modifications of the $p_{\bot}$ spectra due to
final state interactions were obtained in an earlier calculation
\cite{Bass:2000ib}, where the
UrQMD model was used for the final state of a hydrodynamical
calculation with an {\em isothermal} freeze out. 

In both calculations, ours as well as that of Ref.~\cite{Bass:2000ib},
only the $\Omega$ does not change noticeably during the afterburner phase 
because of it's small hadronic cross section. Furthermore there are no
resonance states in the model that contribute to the $\Omega$
spectrum. 

Resonances change the momentum spectra mainly at low momenta, because of the
restricted phase space of the decays and their effects should become
negligible at some point. 
This is demonstrated in Figure \ref{5} where we show the
ratio of the transverse mass spectra after the
resonance decays to those before the decays for four different particles.
The strongest modification of the spectra is observed for the pions
and protons: they are enhanced by a factor of three at low
momenta. However, we find that the effect of resonances is
still considerable even for  $m_{\bot}-m_0 > 1$ GeV, which is
consistent with previous works \cite{Sollfrank:1990qz,Brown:1991en}. Therefore, it
is questionable that an exponential fit below a transverse mass of $m_{\bot}-m_0 < 1$ GeV is
in any way justified, and even at higher transverse masses one would expect small deviations from 
the exponential form.\\

\begin{figure}[t]
 \centering
\includegraphics[width=0.5\textwidth]{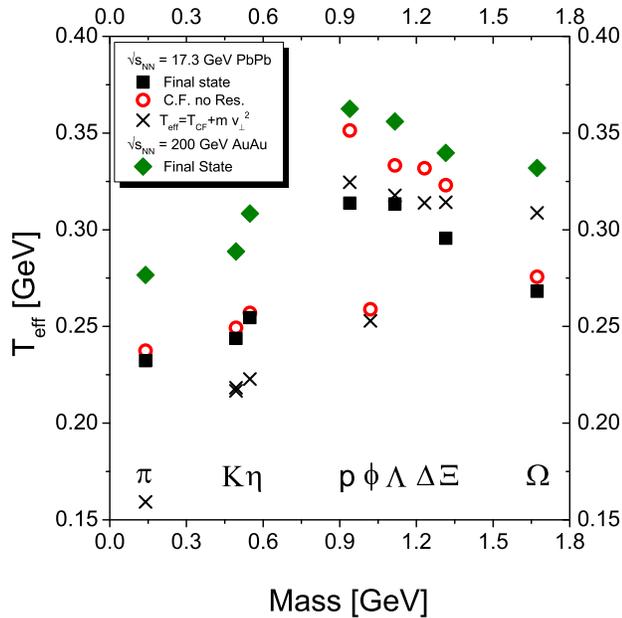}
 \caption{(Color online) Extracted slope parameters $T_{eff}$ from the hybrid model for different particle species. We compare final state results (full symbols) with results obtained at the cooper Frye transition without resonance decays (open symbols).}
 \label{6}
\end{figure}

In Fig.~\ref{6} we show the exponential fits to spectra obtained
with the hybrid model over a mass range of
${1<m_{\bot}-m_0<2}$ GeV.  The
filled symbols correspond to fits of final state spectra, including
resonance decays and the UrQMD afterburner, for most central PbPb/AuAu
collisions at $\sqrt{(s_{NN})} = 17.3\,\rm{GeV}$ (black squares) and $\sqrt{(s_{NN})} =200\,\rm{GeV}$
(green diamonds) GeV. 
The red open circles correspond to fits to
spectra directly after the Cooper Frye transition without resonance decays. One clearly observes a non-monotonic
behavior in the effective slope parameter as a function of the
particles mass. It was shown in earlier publications
\cite{Bass:2000ib} that such a behavior can also be accounted for when
UrQMD is used as the afterburner for an iso-thermal transition. 
In that work the scattering processes in the transport phase create a
situation where the
actual dynamical decoupling takes place over a range of
densities/temperatures leading to results comparable with our study. 
As indicated before, the hydrodynamical and transport
description give equivalent results over a certain range of densities
and our non-isothermal transition is generally not ruled out.
The question whether there is a sharp transition, as in the iso-thermal case,
 or a certain transition region may be further investigated by carefully studying the spectra of
the $\phi$ and $\Omega$, which emerge at the transition from the collective phase to the hadronic phase and do not re-scatter significantly in the transport phase. Their final state observables, therefore, contain the information from the transition surface, without interference from the hadronic phase of the collision. 

To estimate how well the approximation $T_{eff}= T_{th} + m_i \left<
  v_{\bot}\right>^2$, with $\left<v_{\bot}\right>$ taken from figure \ref{3},
compares with our fit results, we plotted the simple estimates of $T_{eff}$ as black crosses. They should be compared with the open red circles because we estimated the flow effect on $T_{eff}$ at the Cooper Frye transition where $T_{th}=T_{CF}$, as we cannot explicitely extract $T_{th}$ from our calculation. It is obvious that the simple formula fails to describe the slope parameter for most of the light mesons and works more reliably for the heavy baryons, indicating that the final hadronic stage is important for the complete description of final particle spectra.     

\section{Elliptic Flow}
The elliptic flow parameter $v_2$ has long been proposed as a probe for the properties of the hot and dense system created in a heavy ion collisions  \cite{Sorge:1998mk,Ollitrault:1992bk,Hung:1994eq,Rischke:1996nq,Sorge:1996pc,Heiselberg:1998es}
\cite{Csernai:1999nf,Brachmann:1999xt,Brachmann:1999mp,Zhang:1999rs,Bleicher:2000sx}. 
The elliptic flow parameter $v_{2}$ is defined as the coefficient of
the second Fourier component of the azimuthal distribution of the emitted particles:
\begin{widetext}
\begin{eqnarray}
\frac{dN}{p_{\bot}dp_{\bot}d\Phi} & = & \frac{1}{2\pi}\,\frac{dN}{p_{\bot}dp_{\bot}}\,[1+2v_{2}(p_{\bot})\cos(2\Phi)+2v_{4}(p_{\bot})\cos(4\Phi)+\ldots]
\label{eq:v2_def_orig}\end{eqnarray}
\end{widetext}
where the azimuthal angle $\Phi=\phi-\psi_{RP}$ is measured with respect to the direction of the reaction plane, $\psi_{RP}$

The average or integrated elliptic flow coefficient, $\bar{v}_{2}$, is defined as:
\begin{equation}
\bar{v}_{2} = \left< cos[2(\Phi)] \right>	
\end{equation}
and is given in terms of the azimuthal distribution as
\begin{eqnarray}
\label{avv1}\bar{v}_{2} & = & \frac{\int dp_{\bot}\,\int_{0}^{2\pi}d\Phi\,\frac{dN}{dp_{\bot}d\Phi}\,\cos(2\Phi)}{\int dp_{\bot}\,\int_{0}^{2\pi}d\Phi\,\frac{dN}{dp_{\bot}d\Phi}}\\ \label{avv2}
&=&\frac{1}{N}\int dp_{\bot}\,\int_{0}^{2\pi}d\Phi\,\frac{dN}{dp_{\bot}d\Phi}\,\frac{p{}_{x}^{2}-p_{y}^{2}}{p_{\bot}^{2}}\\
 & = & \frac{1}{N}\ave{\frac{p_{x}^{2}-p_{y}^{2}}{p_{x}^{2}+p_{y}^{2}}}.
\end{eqnarray}
The integrated elliptic flow is related to the $p_{\bot}$ dependent
elliptic flow by
\begin{equation}
\bar{v}_{2}=\frac{\int dp_{\bot} v_2(p_{\bot}) \frac{dN}{dp_{\bot}}}{\int dp_{\bot} \frac{dN}{dp_{\bot}}}.
\label{intp}\end{equation}

To calculate $\bar{v}_{2}$ in fluid dynamics one usually performs a
Cooper-Frye freeze out and then uses the above
definitions.

Alternatively one can use the energy momentum tensor $T_{\mu\nu}$ and directly extract the flow anisotropy from the hydrodynamical computation \cite{Kolb:1999it,Ollitrault:1997vz} :
\begin{eqnarray}
\bar{v}_{2,T}&=&\frac{\ave{T_{xx}-T_{yy}}}{\ave{T_{xx}+T_{yy}}}\label{rom} \\
&=&\frac{\int d^{3}r\int d^{3}p\frac{\left(p_{x}^{2}-p_{y}^{2}\right)}{p_0}f(r,p)}{\int d^{3}r\int d^{3}p\frac{\left(p_{x}^{2}+p_{y}^{2}\right)}{p_0}f(r,p)}
\end{eqnarray}
Here, $f(r,p)$ is the phase space distribution, and $T_{xx},T_{yy}$
are the appropriate components of the energy momentum tensor. 
This prescription is different from the actual definition of
$\bar{v}_{2}$ which is given by Eq.(\ref{avv2}). It has an additional weighting of $p_{\bot}^{2}/p_0$ in front of $\frac{dN}{dp_{\bot}d\Phi}$.
However, it was empirically found \cite{Kolb:1999it} that the pion
elliptic flow 
can be related to $\bar{v}_{2,T}$ as $2 \ \bar{v}_{2} \approx \bar{v}_{2,T}$. 
Assuming that  $v_2(p_{\bot})\propto p_{\bot}$ and that $f(r,p)$ is
represented by a Boltzmann distribution on can easily show that
$\frac{3}{2} \ \bar{v}_{2} \approx \bar{v}_{2,T}$. However, the
distribution function for pions is considerably distorted by resonance
decays and far from being of Boltzmann type, especially at small
transverse momenta. This brings the factor relating the two definitions
close to the observed value of 2.\\ 

\begin{figure}[t]
 \centering
\includegraphics[width=0.5\textwidth]{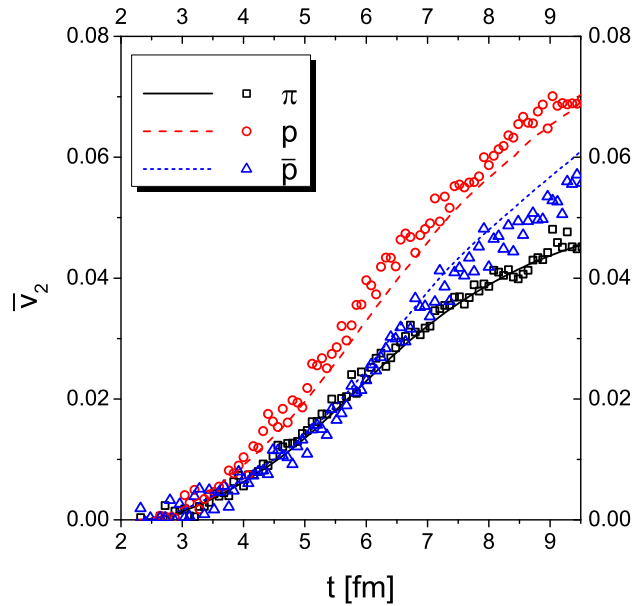}
 \caption{(Color online) Elliptic flow as a function of time for an
   averaged UrQMD initial condition of two Au nuclei colliding at
   $\sqrt(s_{NN})=11.5 A$ GeV and an impact parameter of $b=8$ fm. The
   lines correspond to $\bar{v}_2$ extracted using the energy momentum
   tensor Eq.\ref{rom}, and the symbols denote results from a sampling of the Cooper Frye equation.}
 \label{7}
\end{figure}

\begin{figure}[t]
 \centering
\includegraphics[width=0.5\textwidth]{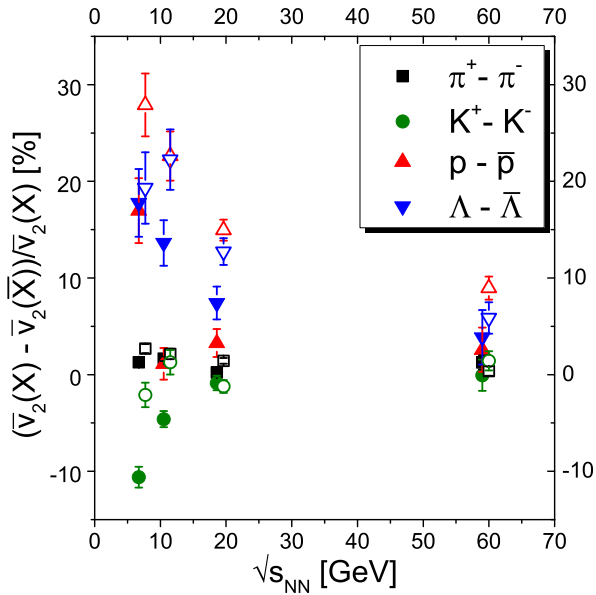}
 \caption{(Color online) Difference of particle and anti particle $\bar{v}_2$ as a function of beam energy for different particle species. The open symbols denote results at the Cooper-Frye transition while the full symbols represent results with the full UrQMD afterburner stage.}
 \label{8}
\end{figure}

In figure \ref{7} we show results for the integrated $\bar{v}_{2}$ of
different particle species, pions protons and anti-protons, as a
function of time, extracted from the hydrodynamical phase of the
hybrid model. For this calculation we used a non-fluctuating
initial condition created by averaging 1000 UrQMD events of collisions
of Au nuclei at an $\sqrt{s_{NN}}= 11.5$ GeV and an impact parameter
of $b=8 fm$. The symbols denote values of $\bar{v}_{2}$ extracted from
sampling the Cooper-Frye equation on an isochronous hypersurface at
each time step, hence representing the correct definition of
$\bar{v}_{2}$. The lines represent values of $\bar{v}_{2,T}$ extracted
from the hydrodynamical energy momentum tensor as described in
equation (\ref{rom}). For the pion elliptic flow we used the full energy momentum tensor and multiplied
$\bar{v}_{2,T}$ by a factor of $0.5$ as suggested in \cite{Kolb:1999it} and one can observe a
very good agreement of the methods. To extract the proton and
anti-proton flow we used the partial energy momentum tensor of the
protons and anti-protons and multiplied them by $2/3$ as suggested above.
The partial $T_{\mu \nu}$'s for protons and anti-protons are related simply by:
\begin{equation}\label{weight}
T_{\mu \nu}^{\bar{p}}=\exp{(-2 \mu_B(x,y)/T(x,y))} \ \cdot \ T_{\mu \nu}^{p} 
\end{equation}
 ($\mu_B$ and $T$ being the baryon chemical potential and temperature respectively).\\
The resulting proton and anti-proton $\bar{v}_2$'s are in reasonable qualitative agreement for both methods presented.\\

In either case we observe that the integrated elliptic flow of protons
is systematically larger than that of the
anti-protons. This difference is also observed in recent
preliminary experimental
data from the STAR Collaboration \cite{Schmah:2011zz}. Below a
collision energy of $\sqrt{s_{NN}} < 60$~GeV the measured elliptic
flow of particles is considerably different from that of their
anti-particles and the difference increases with decreasing beam energy
and correspondingly increasing net baryon density.
A recent transport calculation was able to explain such an effect  qualitatively by the inclusion of mean field type nuclear
potentials \cite{Xu:2012gf}. 
Our calculations, however, indicate that the average flow of
e.g. anti-protons is different from that of protons simply because
their local 'weight', given by $\exp{(-2 \mu_B(x,y)/T(x,y))}$ in the
evaluation of e.g. equation (\ref{weight}), varies due to the finite
net baryon density. As shown in Fig.~\ref{12} the ratio of $\bar{p}/p$ is large for
the cells in the center of the collision which have a high temperature
and small flow velocity, and it is small for the colder cell at the
surface, which carry high flow velocities. As a result the average
transverse flow for protons is larger than that for anti-protons
resulting in a larger value for $\bar{v}_2$ of protons.

\begin{figure}[t]
 \centering
\includegraphics[width=0.5\textwidth]{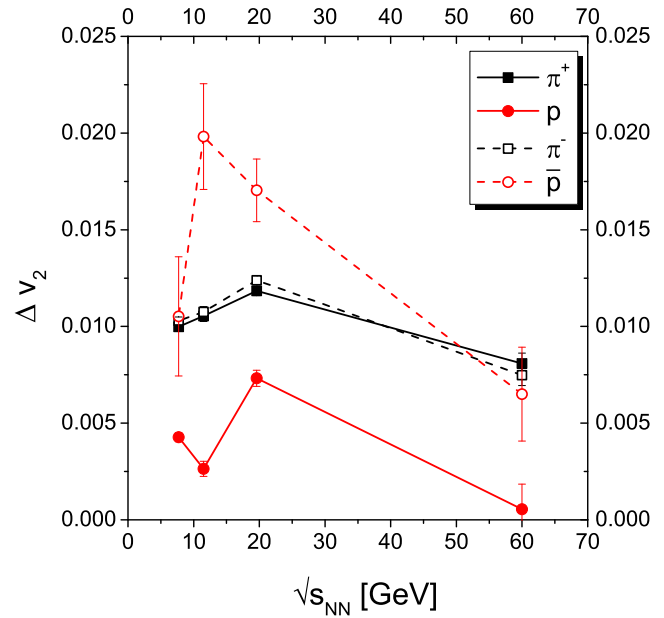}
 \caption{(Color online) Effect of the final hadronic state on particle and anti-particle $\bar{v}_2$ as a function of beam energy. $\Delta v_2$ is defined as the gain of the elliptic flow coefficient obtained in the final transport phase: $\bar{v}_{2}^{Final}-\bar{v}_{2}^{Cooper-Frye}$.}
 \label{10}
\end{figure}

Let us next quantify the discussed difference in elliptic
flow and investigate to which extent our result is modified 
when we apply the full
hybrid model including event-by-event fluctuations, resonance decays
as well as an afterburner stage. Figure \ref{8} shows the difference
of particle$-$anti-particle $\bar{v}_2$ of different particle species as a
function of $\sqrt{s_{NN}}$ for collisions of
Au+Au nuclei at an impact parameter of $b=8$ fm. In the plot the full
symbols correspond to results obtained after the UrQMD final state while the
open symbols denote results after the Cooper-Frye transition,
including resonance decays. In general the difference in elliptic
flow is most pronounced directly after the Cooper Frye transition out
of the hydrodynamical evolution and is washed out by the subsequent UrQMD transport phase.
For baryons we observe a considerable increase in the difference of $\bar{v}_2$ between particles and anti-particles with
decreasing beam energy. The value of $\bar{v}_2$ is essentially the same for $\pi^+$
and $\pi^-$. For the Kaons we observe a trend, which is opposite to
the one found in data \cite{Schmah:2011zz}, showing that $K^-$ $\bar{v}_2$ seems larger than that
of $K^+$. However, this is only the case after the final UrQMD
transport stage. Directly after the hydrodynamical evolution kaons
appear to have the same flow. The observed phenomenon can therefore be interpreted as a natural result of a net-density and a chemically equilibrated phase.

\begin{figure}[t]
 \centering
\includegraphics[width=0.5\textwidth]{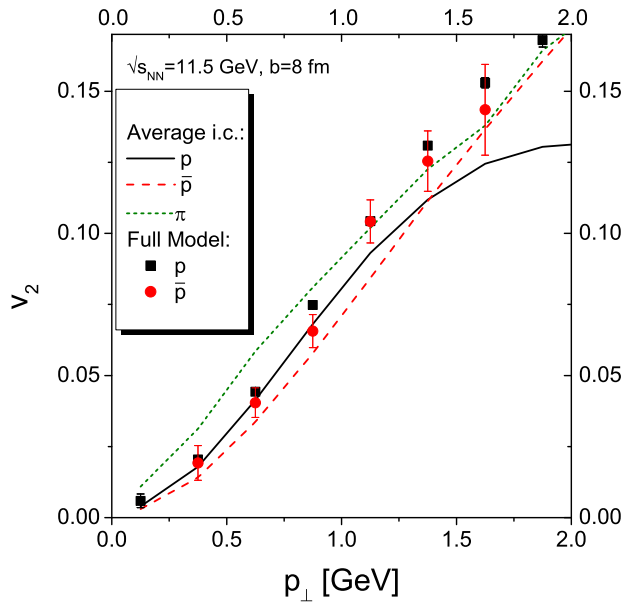}
 \caption{(Color online) Elliptic flow of protons, anti-protons and
   pions as a function of $p_{\bot}$. We show results from an averaged
   UrQMD initial state (lines) as well as from the full hybrid model
   calculations after the Cooper Frye transition (No
     final state interaction) (symbols).}
 \label{9}
\end{figure}

Many transport descriptions fail to even qualitatively describe the
phenomenon because they lack certain interaction channels which are important
at the energies considered here. For example the difference in the Kaon elliptic flow could be explained by missing strangeness exchange reactions in the final state.
For the anti-particles the inclusion of all pair creation processes is
important. Since anti-particles are very rare in
low energy collisions, their flow is sensitive
to their explicit interactions, i.e. annihilation and
recreation, in the transport phase. 
At the highest beam energies a large number of anti-particles is produced at hadronization and regeneration has only a small effect on their abundance (seen in LHC data \cite{sqm,Steinheimer:2012rd}).  
However when the beam energy is decreased,
only few anti-particles are produced and the pair creation due to
multi pion reactions becomes non negligible for the anti-particles bulk properties
\cite{Cassing:2001ds,Rapp:2000gy,Greiner:2000tu,Rapp:2002fc}.\\

To illustrate the effect of the hadronic transport phase on the elliptic
  flow, in Figure~\ref{10} we show the difference between the value
  of $\bar{v}_2$ in the final state to that obtained right after the Copper
  Frye transition for protons, anti-protons and charged
  pions. Clearly, at low energies the value for $\bar{v}_2$ for anti-protons increases
  appreciably during the hadronic transport phase. This is simply an effect of
  the annihilation process. Anti-protons moving in the out-of-plane
  (y) direction encounter more protons to annihilate with than those
  moving in the in-plane (x) direction. Since there are many more
  protons than anti-protons, the annihilation only affects the anti-protons. 
Because the reverse process $n \pi \rightarrow p+\bar{p} $ is not
included in the transport model it is not clear how meaningful the
final state effects (in the UrQMD phase) are for the anti-particle
$\bar{v}_2$.\\

Note that in our calculation the difference between proton and
anti-proton elliptic flow is caused solely by the
non-zero net baryon number density, and chemical potential, and therefore a similar effects
should be be observed for a finite isospin and net
strangeness chemical potential at mid rapidity \cite{Steinheimer:2008hr} if included properly in the model. As discussed at the end of section \ref{s2} the difference in the (elliptic) flow of particles and anti-particles is due to the different local weighting $\exp{(-2 \mu_i(x,y)/T(x,y))}$ when evaluated over the hypersurface. We can assume that the chemical potentials $\mu_i$ have a different magnitude $-\mu_I<\mu_S<\mu_B$ and therefore we expect the effect becomes weaker for Kaons and even more for pions. 
This is in fact the case for the experimental data, as $\pi^-$ show
more $\bar{v}_2$ than $\pi^+$ and $K^+$ more than $K^-$, where the
difference is smallest for the pions. Since the isospin chemical
potential is negative due to the higher abundance of neutron in the
colliding nuclei, this also explains why, contrary to the protons, in
case of the pions the elliptic flow of the anti-particle, i.e. $\pi^-$, is larger than that of the
particle ($\pi^+$).
\\

In figure \ref{3} we see that, during the hydrodynamical evolution, the average transverse momentum of
$\bar{p}$ is smaller than that of $p$.
Therefore, part of
the difference in the integrated $\bar{v}_2$ can be understood as weighting 
$v_2(p_{\bot})$ at higher values of $p_{\bot}$ in case of the protons
(see Eq.(\ref{intp})). On the other hand
the preliminary data by STAR indicate that also the differential
elliptic flow parameter, $v_2(p_{\bot})$, differs between protons and
anti-protons. To investigate this in Fig.~\ref{9} we compare  the
differential elliptic flow for protons, anti-protons and pions after the Cooper-Frye transition. 
The lines represent results with the averaged initial conditions as
described above, where we sampled the freeze out hypersurface to
obtain sufficient statistics. We indeed observe a small
difference for $v_2(p_{\bot})$ between $\bar{p}$ and $p$ . From the
event-by-event calculations we obtain results depicted as
symbols. Here we also only observe a small difference, within the
still considerable errors. In any case, in our calculation we find the
effect due to the different  $\left< p_{\bot} \right>$ to be dominant.\\

\begin{figure}[t]
 \centering
\includegraphics[width=0.5\textwidth]{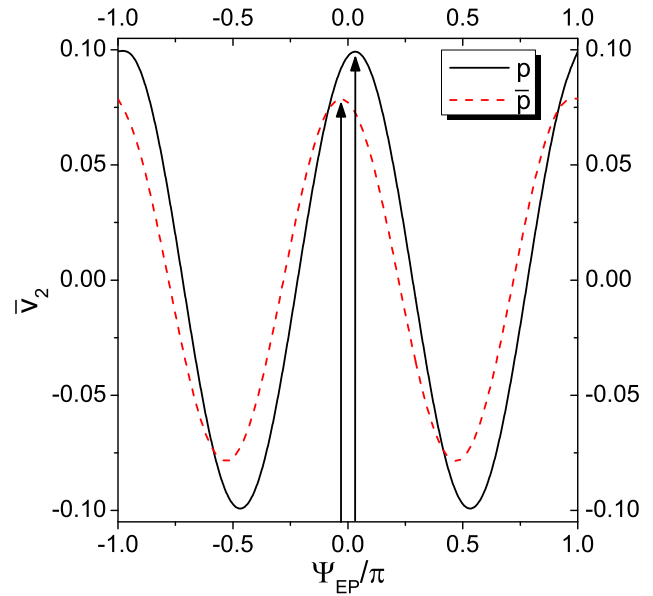}
 \caption{(Color online) Integrated elliptic flow of protons and anti-protons for any given definition of the Event Plane angle $\Psi_{EP}$.}
 \label{11}
\end{figure}

Another effect which can artificially increase the observed asymmetry
of baryon vs. anti-baryon elliptic flow, and in particular the difference in
$p_{\bot}$ dependent $v_2$, may arise from the way $v_2$ is measured
in experiment. 
While in our calculation the reaction plane is well
defined, experiments have to reconstruct the so called Event Plane in
order to infer the Reaction Plane angle $\Psi_{RP}$. This analysis method
involves correlating the azimuthal angle $\phi$ of each particle with
an event plane angle $\Psi_{EP}$ which is defined by the other
particles of that event \cite{Ackermann:2000tr}. Due to fluctuations, 
the Event Plane is usually not perfectly
aligned with the Reaction Plane (see e.g. \cite{Petersen:2010cw}).

Furthermore, as a result of baryon number fluctuations an event plane
direction, $\Psi_{EP}^{[p]}$, 
defined only by protons may deviate from that defined solely
by anti-protons,  $\Psi_{EP}^{[\bar{p}]}$. As a consequence, the elliptic flow for
anti-protons calculated with respect to the event plane of the
protons would be slightly smaller than if it were calculated with respect to
the event plane of the anti-protons. To illustrate this we have
plotted in Fig.~\ref{11} 
the integrated $\bar{v}_2$ for protons and
anti-protons as a function of a trial event plane angle
$\Psi_{EP}$\footnote{ The curves are obtained by calculating one event and evaluating
$\bar{v}_2=\langle cos[2(\phi-\Psi_{EP})]\rangle $ for values of 
$-\pi < \psi_{EP}<\pi $. }. 
Obviously, for $\Psi_{EP}=\Psi_{EP}^{[p]}$ and
$\Psi_{EP}=\Psi_{EP}^{[\bar{p}]}$ the values of $\bar{v}_2$ are
maximal for protons and anti-protons, respectively, indicating the
correct direction of the respective event planes. 
Furthermore, one can clearly see that the maxima for protons and anti-protons are
separated by a finite angle. 

Since at low collision energies  protons are much more abundant than
anti-protons, it is likely that the determination of the event plane
is biased towards the direction of the proton Event Plane, $\Psi_{EP}^{[p]}$.
As a consequence, in each event the value extracted for $v_2$ of {\em
  anti-protons } will be
less than the maximum, as shown in Fig.~\ref{9}. Thus, the event
averaged value for the integrated elliptic flow for anti-protons will 
be systematically smaller than that of protons, even if their true
values would be the same.

At (nearly) vanishing net baryon densities this is not of importance
because one has as many baryons as anti-baryons and there is no bias towards either the proton or anti-proton event plane.
In the case of baryon anti-baryon asymmetry however, the
Event Plane definition will always be biased towards the particle plane.
In the energy range considered we estimated this effect to contribute to the
difference of the flow by about $5-10$ $\%$, by averaging the relative difference of the peak positions over many events.
Note however, that this effect could be excluded if the Event Plane/Reaction Plane can be measured independently, e.g. from the spectator fragments, thus eliminating any bias.

\section{Summary}

We have presented results calculated within the URQMD hybrid model for various
observables from heavy ion collisions. The hybrid model is able to
take into account the initial stopping, the explicit propagation
of the baryon number, and the non-equilibrium transport in the
hadronic phase, all of which are essential ingredients for studies of
relativistic heavy ion collisions at large net baryon
densities. 

We find the observed non-monotonic behavior of $T_{eff}$ with
hadron mass to be a direct consequence of the non-equilibrium
transport in the hadronic phase. Observations like 
the centrality dependence of freeze out
parameters and the difference  in particle and anti-particle spectra
and elliptic flow, on the other hand, can, at least partially, be
explained by the conservation of baryon charge. 

We have argued that the observed difference in
the elliptic flow between positively and negatively charged pions
and kaons may likely be a consequence of the conserved net strangeness
and isospin. Thus, future theoretical studies of heavy ion collisions at low
energies should take into account the explicit conservation of these charges in
addition to the baryon number conservation. We further pointed out that a
transport treatment of the final hadronic phase should include 
both proton -- anti-proton annihilation {\em and} production processes
to ensure detailed balance. This is essential to draw firm conclusions
about the physics leading to subtle difference between particle and
anti-particle observables at low collision energies, such as the observed
difference in the elliptic flow.  
Finally, we have pointed out that local fluctuations of the baryon
number may lead to a biased  determination of the event plane
which may result in artificial differences between particle and anti-particle flow observables. 

\section*{Acknowledgments}
The authors thank U. Heinz for fruitful discussions.
This work was supported by BMBF, HGS-hire and the Hessian LOEWE initiative through the Helmholtz International center for FAIR (HIC for FAIR). J.~S. acknowledges support by the Feodor Lynen program of the Alexander von Humboldt foundation. This work was supported by the Office of Nuclear Physics in the US Department of Energy's Office of Science under Contract No. DE-AC02-05CH11231. The computational resources were provided by the LOEWE Frankfurt Center for Scientific Computing (LOEWE-CSC).

\bibliographystyle{apsrev4-1}

\end{document}